\documentclass[12pt]{iopart}

\usepackage{iopams}
\usepackage{graphicx}
\begin{document}

\title[Distinguishing Majorana bound states and Andreev bound states]{Distinguishing Majorana bound states and Andreev bound states with microwave spectra}

\author{Zhen-Tao Zhang}

\address{School of Physics Science and Information Technology, Shandong Key Laboratory of Optical Communication
Science and Technology, Liaocheng University, Liaocheng 252059, People¡¯s Republic of China}
\ead{zhzhentao@163.com}
\vspace{10pt}
\begin{indented}
\item[]
\end{indented}

\begin{abstract}
Majorana fermions are fascinating and not yet confirmed quasiparticles in condensed matter physics. Here we propose using microwave spectra to distinguish Majorana bound states (MBSs) from topological trivial Andreev bound states. By numerically calculating the transmission and Zeeman field dependence of the many-body excitation spectrum of a 1D Josephson junction, we find that the two kinds of bound states have distinct responses to variations in the related parameters. Furthermore, the singular behaviors of the MBSs spectrum could be attributed to the robust fractional Josephson coupling and nonlocality of MBSs. Our results provide a feasible method to verify the existence of MBSs and could accelerate its application to topological quantum computation.
\end{abstract}

%\noindent(Some figures may appear in colour only in the online journal)
%\submitto{\JPCM}

%\maketitle
\section{Introduction}
Majorana fermions (MFs) have been long-sought particles since they were predicted in 1937. This kind of particle is exotic because its antiparticle is itself. To date, there is no clue that any elementary particle belongs to this kind of particle. On the other hand, some quasiparticles emerged in condensed matter system have been predicted to possess the unique property of MFs \cite{Read00,Kitaev01}, named Majorana zero modes. Importantly, Majorana zero modes behave as non-Abelian anyons, and could be used to build topological quantum computer \cite{Ivanov01,Nayak08,Xue13,Aasen2016}. Therefore, exploring the physical realizations of MF in condensed matter systems has brilliant application prospects in quantum information processing. In the seminal work of Kitaev \cite{Kitaev01}, a toy model was devised that a spinless one-dimensional nanowire with superconductivity could transition to the topological phase, in which a pair of MFs locate at the ends of the nanowire. Along this line, Ref. \cite{Lutchyn10,Oreg10} suggested that a spin-orbit-coupled semiconducting nanowire combining with superconductivity and applied magnetic field could enter topological superconductor phase and host MFs. Recently, several groups reported that they have measured remarkable signatures of MFs in 1D systems \cite{Mourik12,Deng12,Das12,Nadj14}. In these experiments the zero-energy mode of the superconducting nanowire was observed using a tunneling spectrum, which is consistent with MBSs. However, the zero mode can be occasionally reproduced by topological trivial Andreev bound states (ABSs). Thus, MFs cannot be confirmed by the presence of a zero mode. Since then, the question of how to distinguish MBSs and ABSs has been actively pursued and not yet settled in experiments.\\
\indent One important signature of MFs is the fractional Josephson effect \cite{Kitaev01,Kwon04}. In an 1D topological superconductor-normal conductor-topological superconductor(TS-N-TS) junction, the Josephson coupling energy comes from interaction of the pair of MFs near the junction, which is proportional to $\cos{\frac{\phi}{2}}$ with $\phi$ be phase difference across the junction. This lead to the $4\pi$ periodicity of the energy of MBSs formed by the two MFs, unlike the $2\pi$ period of ABSs in topological trivial junction. The period doubling phenomenon is called fractional Josephson effect. However, the effect is built on the conservation of MBSs parity, which could be broken by incoherent quasiparticle poisoning or coherent parity flip due to finite effect of the topological superconductor \cite{Jose12,Zhang17}. Even worse, ABSs could also manifest $4\pi$ periodicity with fine-tuned parameters \cite{Sau12,Sau17}. Therefore, merely observing fractional Josephson effect is not a decisive evidence of MFs \cite{Rokhinson12,Wiedenmann16}. On the other hand, nonlocality is considered as a distinct signature of MBSs. Very Recently, there are serval proposals brought up to verify the existence of MBSs by its nonlocality \cite{Sau15,Rubbert16,Prada17,Hell17}. The main tool used in these proposals is the tunneling spectrum. Besides, a number of other signatures of MFs spring out at the time, for instance, the unusual behaviours of critical current \cite{Jose13} and current susceptibility \cite{Trif18} in topological superconducting junctions, Majorana wavefunctions \cite{Pawlak16}, spin polarization \cite{Jeon17} and optical spectrum \cite{Chen16} in self-assembled topological superconductors. Given the mature technique of exciting transitions between many-body bound states in a weak link using microwave \cite{Bretheau13a,Bretheau13b,van17,Hays17}, it is natural to ask whether the microwave spectrum of Josephson junction could tell us the differences between MBSs and ABSs.\\
 \indent It is well known that if the topological superconductor in a TS-N-TS junction is infinite in length, the pair of MFs in the junction constructs two MBSs, ie., even and odd parity state (0 and 1 state), which can not flip to each other due to parity conservation. Because of the Josephson coupling term $\cos{\frac{\phi}{2}}$, the two states are degenerate at the phase bias $\phi=(2k+1)\pi$, named optimal spots. Ref. \cite{Vayryen15,Peng16,Heck17} has studied the characters of the microwave spectrum of MBSs and ABSs under this condition. However, if the superconductor is not much longer than the superconducting coherence length, the additional pair of MFs resides at the far ends of the junction would effectively couple to the MFs at the junction. Consequently, the degeneracy at the optimal spots is lift and the energy spectrum retrieves to that of ABSs. In this case, to draw the line between MBSs and ABSs, we have calculated the many-body spectra of 1D Josephson junction with various normal transmissions of the junction and Zeeman field. The results show that the spectrum of ABSs and MBSs exhibit two rather different trends as tuning the parameters. In weak or zero Zeeman field, the spectrum of ABSs at the optimal spots appears to be dips, and the energy of the spectrum at the dip strongly relies on the transmission. However, if the magnetic field is approaching the critical value of phase transition the dips would convert to peaks. Furthermore, the magnetic field where the conversion occurs depends on the transmission. The spectrum of ABSs with the lower transmission would alter its shape in weaker Zeeman field. Instead, for MBSs the dips in the spectrum at the optimal spots are maintained at any transmission and any Zeeman field which supports the topological phase. The stability of dips can be owed to the robustness of fractional Josephson coupling across the junction. Besides, the energy of the spectrum at the dips is basically independent of the transmission. The insensitivity of the spectrum at the optimal spots to the transmission is derived from the nonlocality of MFs. In other words, the energy spectrum of MBSs can reveal the effects of robust fractional Josephson coupling and nonlocality, which are not possessed by ABSs. Therefore, we can differentiate MBSs from ABS by using microwave to explore the many-body excitation spectrum of the junction.\\
\indent The paper is organized as follows. In Section 2 we introduce the system and its Hamiltonian. In Section 3 the transmission-dependence of the many-body spectrum of MBSs and ABSs without Zeeman field are compared in a phenomenological manner. After that, we calculate numerically the many-body excitation spectra of the system under different transmissions and Zeeman field in Section 4. Besides, we analysis the differences of the spectra in the topological phase and topological trivial phase. The measurement schemes of the excitation spectrum are present and discussed in Section 5. We conclude this paper in Section 6.
\begin{figure}
\includegraphics[scale=0.8]{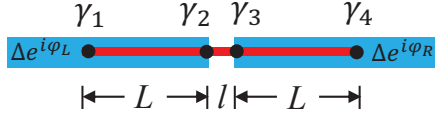}
\centering
\caption{Schematic of SNS Josephson junction. The junction is formed by a spin-orbit coupled nanowire laying on the separate superconductors. If the applied in-plane Zeeman field (not shown) is larger than $B_c$, the two sections of proximitized nanowire could enter topological phase. Four MFs (denoted by solid circles) emerge at the ends of the nanowire. }
\end{figure}
\section{System and Hamiltonian}
The system we consider is a superconductor-normal conductor-superconductor(SNS) junction. The junction consists of a spin-orbit-coupled semiconductor nanowire, which contacts with two separate s-wave superconductors. The Josephson junction geometry is shown in Fig. 1. The left and right sections of the nanowire are superconducting due to proximity effect. For simplicity, we assume that the two superconducting sections of the nanowire have the same length $L$. In addition, we require that the middle section is much shorter than the superconducting coherence length, i.e., $l\ll\xi$, in which case there are only a few subgap bound states localizing at the junction. The Bogoliubov-de Gennes (BdG) Hamiltonian of the nanowire can be written as:
\begin{eqnarray}
\hat{H}=\frac{1}{2}\int dx\hat{\Psi}^\dag(x)H_{NW}\hat{\Psi}(x),\nonumber\\
H_{NW}=(\frac{p_x^2}{2m_*}-\mu-\frac{\alpha_R}{\hbar}\sigma_yp_x)\tau_z+B\sigma_x+\Delta\tau_x,
\end{eqnarray}
where $\hat{\Psi}=(\hat{\psi}_\uparrow,\hat{\psi}_\downarrow,\hat{\psi}^\dag_\downarrow,-\hat{\psi}^\dag_\uparrow)$, and $\sigma_i,\tau_i$ are Pauli matrices
in spin and particle-hole space respectively, $m_*$ is the effective mass, $\alpha_R$ spin-orbit coupling, $B=g\mu_B\mathcal{B}/2$ is the Zeeman splitting resulting from the applied magnetic field $\mathcal{B}$, $g$ is the g-factor and $\mu_B$ the Bohr magneton. Notice that the direction of Zeeman field is perpendicular to the Rashba spin-orbit-coupling field. The induced superconducting gap $\Delta$ reads
\begin{equation*}
\Delta=\cases{\Delta_0e^{i\phi_L}& $x<-\case{l}{2}$\\
0&$|x|\le\case{l}{2}$\\
\Delta_0e^{i\phi_R}& $x>\case{l}{2}$}
\end{equation*}
The superconducting nanowire would transition to topological superconductor phase when the Zeeman splitting exceeds the critical value $B_c=\sqrt{\Delta_0^2+\mu^2}$. Each boundary of the topological superconductor will host a MF. In the system, there are two pairs of MFs: one pair is near the junction, the other locates at the far ends of the nanowire, see Fig. 1. We emphasize that the superconducting nanowire length $L$ are not much longer than coherence length $\xi$, and the finite effect is essential to discriminate MBSs from ABSs.

\section{Transmission dependence of MBSs and ABSs without magnetic field }
In the topological phase or the topological trivial phase in the absence of Zeeman field, the spectrum of the subgap states of the BdG Hamiltonian can be obtained with a low-energy effective theory. The phase-difference dependence of the single-particle energy in the two cases are well-studied. However, what we concern is the trends of the many-body spectrum when varying the transmission of the junction. \\
\indent Let us start with the ABSs spectrum of the SNS junction in the absence of Zeeman field. It is well known that the single-particle energy level is two-fold degenerate, with eigenenergies \cite{Beenakker91}
\begin{eqnarray}
E_\pm=\pm\Delta_0\sqrt{1-T\sin^2{\frac{\phi}{2}}},
\end{eqnarray}
where $T$ denotes the transmission of the junction, $\phi=\phi_R-\phi_L$ is the phase difference across the junction. Due to the double degeneracy, there are four many-body states:$\{|00\rangle,|01\rangle,|10\rangle,|11\rangle\}$. Actually, the states in the odd subspace are hardly populated, therefore we only discuss the even states. It is easy to see that the eigenergies of $|00\rangle$ and $|11\rangle$ are equal to $E_-, E_+$ respectively. \\
\indent Now we turn to the energy spectrum of the junction in the topological phase. In this case, there are four MFs at the ends of topological superconducting nanowire, named $\gamma_1, \gamma_2, \gamma_3, \gamma_4$ (Majorana operators, satisfying $\gamma_i=\gamma_i^\dag$), see Fig. 1. The low-energy effective Hamiltonian can be addressed as
\begin{eqnarray}
H_{eff}=ig_{12}\gamma_1\gamma_2+ig_{23}\gamma_2\gamma_3+ig_{34}\gamma_3\gamma_4
\end{eqnarray}
where $g_{12}(g_{34})$ is coupling strength between $\gamma_1(\gamma_3)$ and $\gamma_2(\gamma_4)$, $g_{23}=\Delta_{e\!f\!f}\sqrt{T}\cos{\frac{\phi}{2}}$ is coupling strength between the middle MFs, $\Delta_{e\!f\!f}$ is effective gap separating MBSs and continuum states of the system in the topological phase. Note that the coupling $g_{23}$ is dependent on the phase difference while $g_{12}, g_{34}$ not. In the effective Hamiltonian, we have neglected the coupling between two next-nearest MFs, such as $\gamma_1$ and $\gamma_3$, or $\gamma_2$ and $\gamma_4$. The reason is that this indirect coupling is much weaker than the interactions between two nearest MFs. To solve the effective Hamiltonian, we introduce the complex fermion operators:
\begin{eqnarray}
a=(\gamma_2+i\gamma_3)/2, \qquad a^\dag=(\gamma_2-i\gamma_3)/2,\nonumber\\
b=(\gamma_1+i\gamma_4)/2, \qquad b^\dag=(\gamma_1-i\gamma_4)/2.\nonumber
\end{eqnarray}
Thus, the low-energy Hilbert space is spanned by the basis $\{|n_an_b\rangle\}$ with $n_a=0, 1$ and $n_b=0, 1$. Because of the conservation of the total parity of the system, the parity-changing transitions are not allowed. Therefore, the Hilbert space can be departed into two decoupled subspaces: even-parity and odd-parity subspace. Without loss of generality, we only investigate the energy eigenstates (ie., MBSs) in the even-parity subspace, $\{|00\rangle,|11\rangle\}$. Diagonalizing the effective Hamiltonian, we can get the many-body eigenenergies:
\begin{eqnarray}
E_{m\pm}=\pm\sqrt{(\Delta_{e\!f\!f}\sqrt{T}\cos{\frac{\phi}{2}})^2+(g_{12}+ g_{34})^2}.
\end{eqnarray}
\indent Comparing Eq. 3 and Eq. 4, we can see that the spectrum in topological phase and topological trivial phase are both $2\pi$ periodic function of phase difference. Moreover, the energy difference between $E_+(E_{m+})$ and $E_-(E_{m-})$ reach their minimum meanwhile, both at the phase difference $(2k+1)\pi$. This means that we can not discriminate them from each other by measuring the phase-difference dependence of energy or its derivation-supercurrent. To overcome this problem, we have studied the transmission dependence of the eigenenergies. Figure. 2 has shown a series of spectra with different transmissions. We can observe that the spectrum in the two phases have explicitly manifested two distinct trends. In topological trivial phase, the positive and negative part of the spectrum would gradually split when we reduce the transmission from its maximum and the gap of the spectrum is expanding accordingly. Contrarily, with the decrease of the transmission, the spectrum in topological phase would shrink in the energy-axis orientation toward the middle of its energy gap, while the energy gap is independent of the transmission. Because at the optimal spots the energy difference between two ABSs or MBSs is prone to be detected with microwave spectrum, in the following we would pay attention to the energy spectrum around these spots. In detail, the energy of the ABS at the optimal spot varies with the transmission while that of the MBS is a constant quantity, which is irrelevant to the transmission of the junction. More interestingly, near the optimal spot, the positive and negative level of the ABSs are lifted when the junction is tuned continuously from the transparent region to the depletion region. Contrarily, the energy of the MBS monotonously descends with the decreasing of the transmission. In a word, the MBS responds to the tuning of the transmission in a rather different style relative to the ABS in the absence of magnetic field. However, the phenomenological results are not capable to distinguish the MBS from the ABS in finite Zeeman field.
\begin{figure}
\centering
\includegraphics[width=9cm,height=5cm]{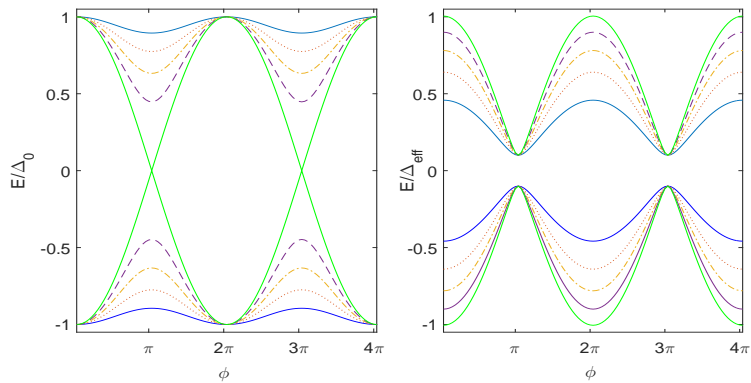}% Here is how to import EPS art
\caption{ Energy spectrum of Andreev Bound states(left) and Majorana bound states(right). The spectrum of ABSs is calculated according to Eq. 2 with different transmissions: $T=0.2$(solid blue line), 0.4(dotted red  line), 0.6(dash-dotted orange line), 0.8(dashed purple line), 1(solid green line). The spectrum of MBSs is obtained from Eq. 3 with different transmissions ( same values as that of ABSs). The Majorana coupling strength is $g_{12}=g_{34}=\Delta_{e\!f\!f}/20$ and is irrelevant with transmission.}
\label{Figure2}
\end{figure}
\section{Many-body excitation spectrum with finite magnetic field: numerical results}
Now we investigate the transmission and Zeeman field dependence of the many-body excitation spectrum of the system. Specifically, our aim in this section is to obtain the energy difference $\Delta\!E$ between the positive and negative many-body eigenstates with even parity ($\Delta\!E$ equal to the twice of the energy of the positive eigenstate due to electron-hole symmetry). To this end, we solve the Hamiltonian in Eq. 1 with a numerical method by applying the tight-binding approximation (for details see the Appendix A in Ref. \cite{Cayao15}). The SNS junction consists of three parts: left and right superconducting part and central normal part. The Hamiltonian is discretized into a tight-binding lattice under the assumption that the parameters in each part take the same values except the gap $\Delta$. The hopping between two nearest-neighbour sites in each part is described by a spin-resolved matrix, named $h$. The interpart coupling is modeled as the hopping between the adjacent sites which lie at the interfaces of the superconducting and normal parts. The interpart hopping matrix $h^\prime$ is related to the intrapart hopping by $h^\prime=\eta h$. In our calculations, we choose the typical values of the parameters for InSb nanowire. The electron's effective mass is $m=0.015m_e$ with $m_e$ be electron's mass, and the spin-orbit coupling strength is $\alpha_R=20meV nm$. For a short junction, it could be proved that the normal transmission rate is vanishing when $\eta\leq0.6$ \cite{Cayao15}, which indicates the tunneling regime. The transparent regime is realized when $\eta\approx1$ in the absence of magnetic field. Therefore, the range $\eta\in[0.6,1]$ is large enough to investigate the energy spectrum on the whole transparency range while reachable in practice.
\begin{figure}
\centering
\includegraphics[scale=0.45]{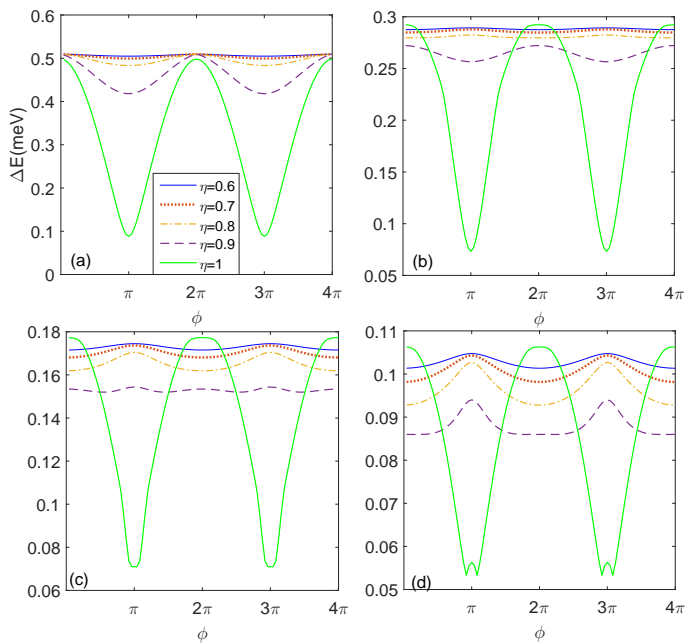}% Here is how to import EPS art
\caption{ Numeric results of many-body excitation spectrum of ABSs. $\Delta\!E$  is calculated by summing the energies of the two lowest ABSs with Zeeman field $B=0$ (a), $0.6B_c$ (b), $0.8B_c$ (c), $0.9B_c$ (d).   In the absence of Zeeman field the spectrum could recover the analytical result shown in Fig. 2. When the magnetic filed is approaching the critical value $B_c$, the spectrum is qualitatively different from that without magnetic field. The parameters are: $\Delta=0.25meV$, $\mu=0.5meV$, $\alpha_R=20meV nm$, $L=2\mu m$, $l=10nm$. }
\label{Figure3}
\end{figure}
\subsection{Many-body excitation spectrum of ABSs}
We have calculated the many-body excitation spectrum with different transmissions when the applied magnetic field is lower than the critical value, see Fig. 3. In the absence of  magnetic field our numerical result Fig.3(a) is consistent with the analytical spectrum shown in Fig. 2. The minimum of the energy in the spectrum is locating at the optimal spots no matter how weak the transparency ($\eta$) of the junction is. However, the situation is changed when the system is subjected to a magnetic field. From Fig.3(b) where $B=0.6B_c$, we see the low $\eta$ spectrum springs up a peak at the optimal spot while the high $\eta$ spectra appear as a dip at the same phase bias. When $B=0.8B_c$, only the spectrum with $\eta=1$ remain a dip at the optimal spot (Fig.3(c)). As up to $B=0.9B_c$, there has been a local peak emerged within the dip of the $\eta=1$ spectrum. Therefore, we can conclude that when approaching the phase transition the spectrum at the phase differences $\phi=(2k+1)\pi$ forms a peak at arbitrary transparency. Moreover, given a certain magnetic filed, we find that the energy at the optimal spot varies remarkably with the transmission. This phenomenon is expected since ABS is localized at the junction and its energy strongly depends on the transmission of the junction.
\begin{figure}
\centering
\includegraphics[scale=0.45]{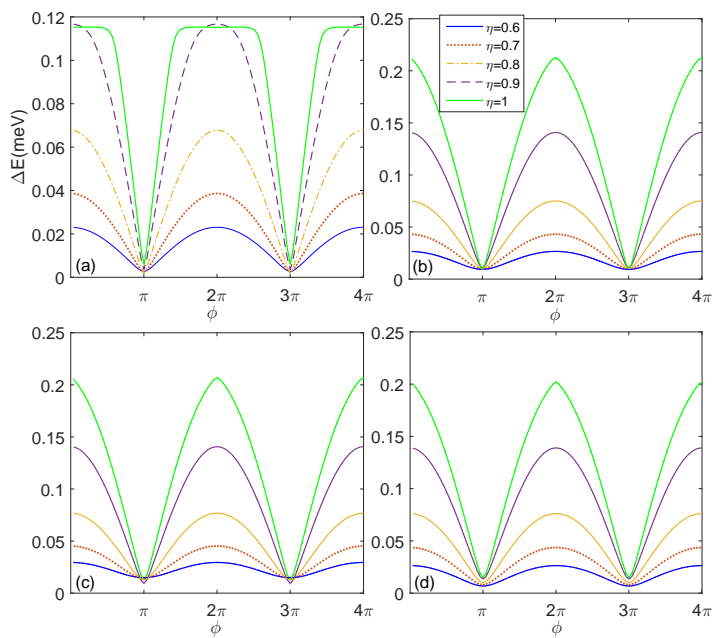}% Here is how to import EPS art
\caption{ Numeric results of many-body excitation spectrum of MBSs.  The Zeeman field is $B=1.2B_c$ (a), $1.4B_c$ (b), $1.6B_c$ (c), $1.8B_c$ (d) respectively. The other parameters are as same as Fig. 3.}
\label{Figure4}
\end{figure}
\subsection{Many-body excitation spectrum of MBSs}
Once the applied Zeeman field exceeds the its critical value, the local ABSs would evolve into nonlocal MBSs. The many-body excitation spectrum in this regime is illustrated in Fig. 4. We can see that MBSs behave quite differently compared with ABSs when the transmission and Zeeman field are tuned. Firstly, the many-body spectrum of MBSs always forms a dip at the optimal spot. The dip persists in a very wide range of Zeeman field. From Fig. 4(a) to (d) with the Zeeman field changing from $1.2B_c$ to $1.8B_c$, there is no hint that the dips at the optimal spots would evolve into peaks. Moreover, the dips are also robust to the variation of the transmission.  The stability of the dips at the phase difference $\phi=(2k+1)\pi$ reflected the robustness of the term $E_m\cos{\frac{\phi}{2}}\gamma_2\gamma_3$ in the Hamiltonian (Eq. 3) to the Zeeman field and the transmission. The $\cos{\frac{\phi}{2}}$ term resulting from MFs coupling is the source of fractional Josephson effect. In contrast, the dips at the spectrum of ABSs is instable as long as varying the relevant parameters. Secondly, the energy of MBSs at the optimal spots is almost invariant in a fixed magnetic field no matter how large the transmission is. We find that this unusual phenomenon is a direct consequence of the nonlocality of MBSs. As shown in Eq. 4, the energy at the optimal spot within topological phase is determined by the coupling strength $g_{12}$ and $g_{34}$, which describe the interactions between MFs at the same side of the junction. Due to the symmetry of the junction, $g_{12}$ and $g_{34}$ hold the same relation with the transmission, therefore we only investigate the former. $g_{12}$ is proportional to the overlap of the wavefunction of $\gamma_1$ and $\gamma_2$. Under the assumption that the superconducting nanowire length $L$ is larger than the coherence length $\xi$, the wavefunction overlap takes place in the topological section of the nanowire. Furthermore, the overlap relies on the ratio between $L$ and $\xi$, both of which are theoretically independent of the transmission of the junction. In other words, the MBSs distributed within the topological nanowire is insensitive to a local perturbation outside the topological nanowire. This contrasts sharply with the situation of ABSs which is localized at the junction. Therefore, the invariance of the spectrum at the optimal spots can be considered a convincing signature of the nonlocality of MBSs.
\section{Measurement schemes }
\indent Before concluding this paper, we would like to discuss how to probe the excitation spectrum of ABSs and MBSs with experimentally reachable techniques. Basically, there are two feasible schemes to achieve this. The first is embedding the nanowire in a hybrid superconducting quantum interference device (SQUID) whose second arm is a conventional tunnel junction with Josephson coupling energy much larger than that of the nanowire-based junction \cite{van17}. In the case of a negligible loop inductance of the SQUID, the applied phase with external magnetic flux mostly drops over the nanowire junction. That means the phase difference $\phi$ could be tuned at will. The microwave emitted by the tunneling junction would be absorbed by the investigated nanowire junction if the microwave frequency match the many-body transition energy shown in the last section. One can measure the microwave response of the nanowire junction utilizing another tunnel junction capacitively coupled to the hybrid SQUID.\\
\indent The second measurement scheme is employing the techniques of circuit quantum electrodynamics (cQED). The main point is that the loop containing a nanowire Josephson junction is inductively coupled to a superconducting microwave resonator. The interaction between the two even-parity ABSs (or MBSs) and the resonator is well-described by Jaynes-Cummings model. In the dispersive regime of cQED, the frequency of the resonator would be shifted by a magnitude that is dependent on the state of the ABSs (or MBSs). Thus, the state of ABSs (or MBSs) could be detected by observing the resonator response to a microwave with a proper frequency. In addition, the odd parity states have no effect on the resonator, therefore could be discriminated from even parity states. With this state-readout approach at hand, we could further probe the excitation spectrum of the nanowire junction by the conventional spectroscopic means.
Very recently, Ref. \cite{van17} and Ref. \cite{Hays17} have performed alike spectral measurements of ABSs in nanowire Josephson junction using the above addressed schemes respectively, though in the absence of or in weak Zeeman field. Therefore, it is hopeful to extend the spectral measurements into strong Zeeman field condition under which the topological phase transition might occur. In experiment, the different styles of the dependence of the excitation spectrum on the transmission would mark out the different topological phases.

\section{Conclusion}
\indent In this paper we have calculated the transmission and Zeeman field dependence of many-body spectrum of a 1D Josephson junction system. We find that the spectrum in topological phase and topological trivial phase have distinct responsibilities to the variations of the related parameters. Moreover, the unusual properties of the MBSs stem from the robustness of the fractional Josephson coupling and nonlocality of the MBSs. Our study provides a feasible method to verify the existence of MBSs and could push forward the MF-based topological quantum computation.
\section*{ACKNOWLEDGMENT}
This work was funded by the National Science Foundation of China (No.11404156), and the
Startup Foundation of Liaocheng University (Grant No.318051325).


\section*{References}
\begin{thebibliography}{31}
\bibitem{Ivanov01}Ivanov D A 2001 {\it Phys. Rev. Lett.} {\bf 86} 268
\bibitem{Nayak08}Nayak C, Simon S H, Stern A, Freedman M and Das Sarma S 2008 {\it Rev. Mod. Phys.} {\bf 80} 1083
\bibitem{Xue13}Xue Z Y, Shao L B, Hu Y, Zhu S L and Wang Z D 2013 {\it Phys. Rev. A} {\bf 88} 024303
\bibitem{Aasen2016}Aasen D {\it et al} 2016 {\it Phys. Rev. X} {\bf 6} 031016
\bibitem{Read00} Read N and Green D 2000 {\it Phys. Rev. B} {\bf 61} 10267
\bibitem{Kitaev01} Kitaev A Y 2001 {\it Phys. Usp.} {\bf 44} 131
\bibitem{Lutchyn10}Lutchyn R M, Sau J D and Das Sarma S 2010 {\it Phys. Rev. Lett.} {\bf 105} 077001
\bibitem{Oreg10}Oreg Y, Refael G and von Oppen F 2010 {\it Phys. Rev. Lett.} {\bf 105} 177002
\bibitem{Mourik12}Mourik V {\it et al} 2012 {\it Science} {\bf336} 1003
\bibitem{Deng12} Deng M T {\it et al} 2012 {\it Nano Lett.} {\bf 12} 6414
\bibitem{Das12} Das A {\it et al} 2012 {\it Nat. Phys.} {\bf 8} 887
\bibitem{Nadj14} S. Nadj-Perge \emph{et al} 2014 {\it Science} {\bf 346} 602-607
\bibitem{Kwon04} Kwon H J, Sengupta K and Yakovenko V M 2004 {\it Eur. Phys. J. B} {\bf 37} 349
\bibitem{Jose12}San-Jose P, Prada E and Aguado R 2012 {\it Phys. Rev. Lett.} {\bf 108} 257001
\bibitem{Zhang17}Zhang Z T, Xue Z Y and Yu Y 2017 {\it Europhys. Lett.} {\bf 118} 57005
\bibitem{Sau12} Sau J D, Berg E and Halperin B I 2012 arXiv:1206.4596
\bibitem{Sau17}Sau J D and Setiawan F 2017 {\it Phys. Rev. B} {\bf 95} 060501
\bibitem{Rokhinson12} Rokhinson L P, Liu X and Furdyna J K 2012 {\it Nat. Phys.} {\bf 8} 795
\bibitem{Wiedenmann16}J. Wiedenmann {\it et al} 2016 {\it Nat. Commun.} {\bf7} 10303
\bibitem{Sau15}Sau J D, Swingle B and Tewari S 2015 {\it Phys. Rev. B} {\bf 92} 020511
\bibitem{Rubbert16} Rubbert S and Akhmerov A R 2016 {\it Phys. Rev. B} {\bf94} 115430
\bibitem{Prada17}Prada E, Aguado R and San-Jose P 2017 {\it Phys. Rev. B} {\bf96} 085418
\bibitem{Hell17}Hell M , Flensberg K and Leijnse M 2017 arXiv:1710.05294
\bibitem{Jose13}San-Jose P, Cayao J, Prada E and Aguado R 2013 {\it New Journal of Physics} {\bf 15} 075019
\bibitem{Trif18}Trif M {\it et al} 2018 {\it Phys. Rev. B} {\bf97} 041415
\bibitem{Pawlak16}Pawlak R {\it et al} 2016 {\it npj Quantum Information} {\bf 2} 16035
\bibitem{Jeon17} Jeon S {\it et al} 2017 {\it Science} {\bf 358} 6364
\bibitem{Chen16} Chen H J {\it et al} 2016 {\it Scientific Reports} {\bf 6} 36600
\bibitem{Bretheau13a}Bretheau L{\it et al} 2013 {\it Nature} {\bf499} 312
\bibitem{Bretheau13b} Bretheau L {\it et al} 2013 {\it Phys. Rev. X} {\bf3} 041034
\bibitem{van17}van Woerkom D J {\it et al} 2017 {\it Nat. Phys.} {\bf13} 876
\bibitem{Hays17} Hays M {\it et al} 2017 arXiv:1711.01645
\bibitem{Vayryen15} V\"{a}yrynen J I, Rastelli G, Belzig W and Glazman L I 2015 {\it Phys. Rev. B} {\bf92} 134508
\bibitem{Peng16}Peng Y, Pientka F, Berg E, Oreg Y and von Oppen F 2016 {\it Phys. Rev. B} {\bf94} 085409
\bibitem{Heck17}van Heck B, V\"{a}yrynen J I and Glazman L I 2017 {\it Phys. Rev. B} {\bf96} 075404
\bibitem{Beenakker91} Beenakker C W J 1991 {\it Phys. Rev. Lett.} {\bf 67} 3836
\bibitem{Cayao15} Cayao J, Prada E, San-Jose P and Aguado R 2015 {\it Phys. Rev. B} {\bf91} 024514

\end{thebibliography}
\end{document}